\documentclass [amssymb, amsmath,pre,nofootinbib,showkeys]{revtex4}
\pdfoutput=1

\usepackage{amsmath,bm}
\usepackage{graphicx}
\usepackage{color}
\usepackage{subfigure}
\usepackage{url}

\begin{document}
\large

\title{Non-Equilibrium Statistical Physics of Currents in Queuing Networks}

\author{Vladimir~Y.~Chernyak $^{a,b}$ }
\email{chernyak@chem.wayne.edu}
\author{Michael~Chertkov $^{a,c}$}
\email{chertkov@lanl.gov}
\author{David~A.~Goldberg $^{a,d}$}
\email{dag3141@mit.edu}
\author{Konstantin Turitsyn $^{a,e}$}
\email{turitsyn@lanl.gov}

\affiliation{
$^a$Center for Nonlinear Studies and Theoretical Division, LANL, Los Alamos, NM 87545\\
$^b$Department of Chemistry, Wayne State University,
5101 Cass Ave,Detroit, MI 48202\\
$^c$New Mexico Consortium, Los Alamos, NM 87544\\
$^d$ Operations Research Center, MIT, Cambridge, MA 02139\\
$^e$ Landau Institute for Theoretical Physics, Moscow, Russia, 119334}

\date{\today}

\begin{abstract}
We consider a stable open queuing network as a steady non-equilibrium system of interacting particles.  The network is completely specified by its underlying graphical structure, type of interaction at each node, and the Markovian transition rates between nodes. For such systems, we ask the question ``What is the most likely way for large currents to accumulate over time in a network ?'', where time is large compared to the system correlation time scale. We identify two interesting regimes.  In the first regime, in which the accumulation of currents over time exceeds the expected value by a small to moderate amount (moderate large deviation), we find that the large-deviation distribution of currents is universal (independent of the interaction details), and there is no long-time and averaged over time accumulation of particles (condensation) at any nodes.  In the second regime, in which the accumulation of currents over time exceeds the expected value by a large amount (severe large deviation), we find that the large-deviation current distribution is sensitive to interaction details, and there is a long-time accumulation of particles (condensation) at some nodes.  The transition between the two regimes can be described as a dynamical second order phase transition.  We illustrate these ideas using the simple, yet non-trivial, example of a single node with feedback.
\end{abstract}


\maketitle

\section{Introduction}
\subsection{Non-equilibrium statistical physics and queueing networks}
The concept of statistical equilibrium is extremely powerful. Once detailed balance (which is synonymic to the equilibrium) is established, one can shortcut a discussion of dynamics and just consider the Gibbs distribution that governs simultaneous correlations in the steady state. On the other hand, if detailed balance is broken no free lunch is guaranteed, and one generally does need to dive into dynamics, even to describe just the steady state. This is an infamous and principal difficulty at the very core of non-equilibrium statistical physics. It is thus of interest to identify a class of non-equilibrium steady systems where the steady state, distinctly different from the Gibbs distribution, can be derived in a straightforward way.

\begin{figure}
\includegraphics[width=3in,page=3]{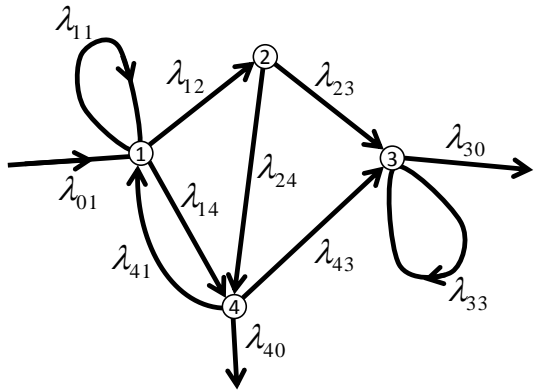}
 \caption{\label{fig:network_example} Example of an open finite queueing network represented by a directed graph. The sample graph consists of four vortexes/stations, labeled $1,2,3,4$, with label $0$ is reserved for an external (out) node.  Transitions between the stations are shown as directed edges. Loops (self-loops), as $1\to 1$, are allowed. Each graph edge is equipped with a transition rate. Throughout the manuscript all transitions in the network are assumed memoryless, i.e. $0$-degree Markovian - Poisson. We also focus on the case of the infinite waiting room, i.e., no particles/jobs are lost, and thus the number of particles accumulated at a node may reach $+\infty$. Another important characteristic of a station/note is the number of tellers/servers. For the sake of simplicity, we assume the jobs/particles to be identical (single-class), and all tellers at any given station processing with the same time-independent Poisson rate.}
 \end{figure}

The problem discussed in this manuscript belongs to this class - it is a non-equilibrium statistical physics problem with the steady state known explicitly. More precisely, we study the model generally known as an open queueing network.  A general (Markovian) open queueing network, the so-called Jackson network \cite{63Jac}, can be described as a random walk of particles (jobs, vehicles, people, computer packets, etc) on a directed graph. The directed links label the transitions between stations/nodes, each of the Poissonian type and thus characterized by a single number (the rate). Each node is characterized by the number of equivalent servers. Collectively this defines a many-particle problem, whose non-equilibrium nature (no detailed balance) follows immediately from the definitions. Adopting the terminology of the Queueing Theory community (a subset of the Operations Research community), the type of service at a node is provided by the $M/M/m/\infty$ queue, which is translated as Markovian input, Markovian output with $m$ servers, and infinite waiting room. (Throughout the manuscript we use the shorter notation $M/M/m$.) An example of such a network is shown in Fig.~\ref{fig:network_example}, and more information will be provided below.  Note that a similar, and to a degree more general, model is known in statistical physics as the zero range model \cite{70Spi}.

In spite of the generally pessimistic non-equilibrium assessment, the stable (i.e. achieving a statistical steady state) open Jackson network allows an explicit and simple solution for the steady state \cite{76Kel}. Similar statements apply to the zero range model \cite{70Spi}. The solution for the steady state is the so-called product form \cite{70Spi,76Kel,N93}, where the joint distribution function of the occupation numbers at all the nodes/stations is factorized into a product of marginal probabilities at the separate nodes. Some new and physics-based exposition of this factorized solution is due to \cite{07Zei}, and will also be a part of our construction below. We note that a different factorization has also led to the solution of the related Asymmetric Exclusion Processes (AEP) Models, see \cite{92DDM,98DL,07Der,07BE} and the references therein \footnote{The general AEP models are significantly different from the Jackson-network models discussed in this manuscript - they can be viewed as a special case of an $M/M/1/1$ queuing model (waiting room with one slot and no particles lost) in contrast with the Jackon network of $M/M/m/\infty$ (infinite waiting room) queues. In the case of a simple one-dimensional chain the AEP model is reducible (with a proper redefinition of the phase space) to the Jackson network model (and other way around) \cite{92DDM,98DL,07Der,07BE}, and then both models show product-state form solution \cite{70Spi,76Kel,N93}. However,  the product-state solvability,  which holds for the Jackson network generally,  does not extend to the AEP model on general graphs.}.

On the other hand, in recent years the quest for universality in non-equilibrium statistical problems has turned to the analysis of currents generated over time, long compared to the system correlation time scale. This route became fruitful and helped to establish some fundamental relations about the symmetry of the currents distribution, known as fluctuation theorems \cite{95GC,98Kur,99LS}. In spite of this partial success, the full description of the currents distribution for a general many-particle and open non-equilibrium problem was (and is) deemed too difficult
\footnote{We would like to emphasize here the principal difficulty arising from the fact that the system is both (a) many-particle and (b) open.  The analysis of such systems is often limited to the setting in which there are few degrees of freedom.  An example is a recent paper of three of us with Puliafito \cite{07TCCP} that discusses an explicit expression for the distribution of currents associated with a polymer stretched by external shear flow. In another paper (two of us with Malinin and Teodorescu \cite{09CCMT}) considered a setting with multiple degrees of freedom and analyzed the statistics of so-called topological currents. The statistics of currents have also been studied for the aforementioned AEP models over one-dimensional chains, which typically have multiple degrees of freedom \cite{07Der}.}.

\subsection{Large deviations, queueing networks, and condensation phenomenon}\label{qnetflows}
One of the central questions in the Queueing Theory community is the probability of rare events in queueing networks (large deviations) \cite{01CY}.  A central tool in understanding such events
is the notion of the so-called `fluid-limit' of a queueing network \cite{CM91}, which was originally developed to understand certain questions related to the stability (existence of steady state) for certain complicated networks \cite{D95,RS92}.  We now describe this fluid-limit in greater detail.  For a \emph{fixed} queueing network, one scales both \textbf{time} and the \textbf{initial number of particles in system} by some large integer $n$, and then normalizes the set of queue lengths by $n$.  Note the important difference with the standard notion of hydrodynamic limit - here, the number of nodes in the network is \textbf{fixed} - only time and the initial number of particles are scaled.  It is proven in \cite{CM91,01CY} that for a broad class of networks, this scaling has a non-trivial limit (the fluid limit).  It is then standard in the Queueing Theory literature to perform the large deviations analysis in the setting of this fluid limit, as opposed to the original (unscaled) network.  There has been great progress in understanding all aspects of these large deviations \cite{AD99,Ananth90,00Ign,08MR,07Puh}. However, driven by applications to describing buffer overloads in communications systems, most of these results have been geared towards understanding how large queue lengths accumulate over time (see for example \cite{08MR}).  Much less is known about how large currents build up in a queueing network over time, especially for the unscaled (not fluid limit) network.
\\\indent  The statistical physics community has developed several tools ( e.g. the fluctuation theorems \cite{95GC,98Kur,99LS} ) that are well-suited to studying the large-deviations properties of currents in a network.  In light of the aforementioned gap in our understanding of the large deviations of currents in queueing networks and related network models, several researchers in the statistical physics community have recently begun to apply the fluctuation theorems to the study of currents in a variety of network-related models \cite{Phys1MK10,Phys2RH08,Phys3HRS06,Phys4HRS05}.
Most of these results have been for models such as the zero-range process, which are tangential (although closely related) to the networks studied by the Queueing Theory community.  We bridge this gap by directly applying such an analysis to the canonical model of queueing theory - namely the Jackson network.  We note that although the potential application of these tools to Jackson networks is mentioned in \cite{Phys1MK10}, the paper closest in spirit to our own is \cite{Phys2RH08}.  Indeed, for the 1-D zero-range process the authors identify several regimes in which the model may operate, characterized by the large-deviations properties of currents, and whether or not there is an accumulation of particles at sites over time (so-called condensation phenomenon).  Their analysis proceeds by formalizing the system dynamics in a quantum-mechanical / operator-theoretic framework, and then studying the relevant spectral properties and Cr\'{a}mer (large deviation) functions.
Our own analysis will very-much parallel that of \cite{Phys2RH08}, but for the Jackson network model on general graphs.  Other related work can be found in \cite{06BDGJL}, in which the zero-range model on regular lattices is studied in the hydrodynamic limit, as well as \cite{02BDGJL} where the closely related stochastic lattice gas model is studied.
\\\indent We also note that the Queueing Theory community has already had some success in applying some of these tools \cite{S93,MY96,Stolyar09}.  Indeed, these analyses (and our own) rely on the interpretation of queueing systems as interacting particle systems, which (historically) helped lead to many of the great breakthroughs in the understanding of queueing networks (e.g. product-form solution \cite{79Kel}).  However, these analyses have been for either an infinite 1-D chain of queues \cite{S93,Stolyar09} or only in the hydrodynamic limit \cite{MY96}.  In either case, their findings parallel our own, in the existence of certain phase transitions in which condensation may (or may not) occur.

\subsection{Currents in a queueing networks with feedback}\label{feedbackground}
We now discuss the nature of the currents in a Jackson network in steady state, a topic that has generated much research in the queueing literature \cite{B56,BM78,PS79,LPS81,81WV}.
Although currents over the entering links are Poissonian by construction, those which leave the system are not obviously Poissonian.  However, if the system is stable, these exiting flows are in fact Poissonian \cite{BM78,79Kel}. More generally, the flow is Poissonian along all arcs that may not be revisited by a particle, and these flows enjoy several nice properties (such as asymptotic independence) \cite{81WV}.  It was recognized early on in the queueing literature that the statistics of the internal currents with feedback are rather difficult to analyze \cite{BM78}.
The source of these difficulties is the complicated feedback mechanism that arises when particles may revisit an arc.  Thus, even for the simple single-node feedback system (which will serve as an enabling example in this manuscript), the statistics of the feedback current are rather complicated \cite{76Bur,78Bre,78BM}.  Although several properties of the currents in a multi-server feedback queue and its generalizations have been studied in the literature \cite{63Tak,02PJ,91DKW,80DMS,77DD,75DD}, most of these results involve showing that in certain limiting regimes the flows are close to Poissonian under some metric \cite{02PJ,05BFX,00BWX,96BB} - the precise nature of these flows remains poorly understood.  Furthermore, it seems that an operator-theory/generating-function approach has not yet been applied to the study of these feedback currents.  We will take this approach to derive new understanding of these currents.

\subsection{\underline{\bf Our main results} }
The main results of our manuscript can be described as follows:
\begin{itemize}
\item We present an explicit, detailed operator-theoretic description of a Jackson network in the Doi-Peliti formalism.  This expands on the description given in \cite{07Zei}, with an eye towards  introducing a larger set of the statistical physics community to standard queueing models.
\item We identify an ``uncongested" regime in Jackson networks, w.r.t. the large-deviation behavior of current.  This regime is universal, i.e. interaction-independent, and non-Poissonian.  Furthermore, in this regime there is no infinite accumulation of particles at any node (condensation).  This universality can be qualitatively explained as follows. In this regime, the given deviation is driven by sample paths in which the number of particles does not diverge.  This will generally occur when the given deviations are somewhat mild, and a deviation can be attained without a massive buildup of particles. Then, the impact of one particle `blocking' another does not contribute asymptotically to the deviation, and thus the system is equivalent asymptotically to one in which all nodes have an infinite number of servers (particles do not interact).  We confirm the existence of this regime by demonstrating that the single-node network with feedback falls into this regime for certain parameters, which we compute explicitly.
\item We also identify a second, ``congested" regime, which is interaction-dependent. In this regime, a given deviation is driven by system primitives that are \emph{dependent} on the number of servers and service times.  This will generally occur when the given deviations are more severe, and the only way to attain the given deviation is to have all, or at least some, servers busy for essentially the entire time horizon.  In this regime the time-averaged queue length diverges for at least one node, marking the dynamical phase transition between the two regimes as second-order.
\item We observe that statistics of time-averaged (over the large observational interval) queue and queue measured at the last moment of time, both conditioned to an atypical (large or small) values of currents, are not identical. The difference is particularly striking in the ``congested" regime of the largest currents, where all moments of the former object (queue at the last moment of time) saturate to finite values while all moments of the later object (time-averaged queue) diverge with time. Quite generally this phenomenon can be classified as a breakdown of ergodicity in cases which are atypical with respect to currents.
\end{itemize}

\subsection{Outline}
The manuscript is organized as follows. In Section \ref{sec:DPM}, we present a technical introduction to the dynamics of the Jackson network in terms of the physics-native Doi-Peliti (``quantum" or ``second-quantized") technique \cite{76Doi,85Pel,86Pel}.  There we formally characterize the Master Equation (ME) that governs the system evolution and steady state, as well as the joint distribution of densities (occupation numbers).  We customize this description to the $M/M/\infty$, $M/M/1$, and general $M/M/m$ models in Sections \ref{subsec:MMinfty}, \ref{subsec:MM1}, and \ref{subsec:MMm}, respectively.  In Section \ref{sec:Currents}, which represents the core of the manuscript, we adopt the Doi-Peliti technique to analyze the ME for the joint distribution function of densities (that reside at the nodes) and currents (that reside at the links).  We also show how the coherent-state technique provides a complete description of the ground state (eigen-value and eigen-function) in
 terms of the relevant evolution operator.  Section \ref{sec:Currents} is partitioned into three Subsections and an Appendix.  In Section \ref{subsec:uncongested}, we discuss the universal (and statistically typical) ``uncongested" regime.  We also describe the boundary of the ``uncongested" region in the space of currents, and comment on the associated dynamical phase transition.
Our analysis procedes by invoking an auxiliary construction for the left eigen-function of the evolution operator, which is discussed in Appendix \ref{app:Left}. The transitions between the ``uncongested'' and (partially) ``congested'' regimes are discussed in Section \ref{subsec:congested}. The general theory is illustrated in Section \ref{subsec:feedback} for a single-node system with feedback. In Section \ref{sec:feedback}, to validate the theory, we describe a full spectral solution for this single-feedback problem. In Section \ref{sec:path} we draw several conclusions and discuss future directions for research.

\section{The Doi-Peliti-Massey operator technique for a Generic Birth-Death Process}
\label{sec:DPM}

This Section introduces notation and describes the main operational rule of the Jackson network in terms of the statistical physics native Doi-Peliti technique \cite{76Doi,85Pel,86Pel}.
We note that a very similar (but less explicit) formulation was derived in \cite{07Zei}.

As we will see, in this context the Doi-Peliti technique is closely related to the operator-theoretic framework formulated by Massey \cite{84Mas_a,84Mas_b} in the Queueing Theory community. We start by introducing the quantum-mechanics based bra(c)ket notation.  We then discuss the product-form solutions for the stationary problems associated with the $M/M/\infty$, $M/M/1$, and generic $M/M/m$ networks in Sections \ref{subsec:MMinfty},\ref{subsec:MM1}, and \ref{subsec:MMm}, respectively.

The network ( e.g. the one shown in Fig.~(\ref{fig:network_example}) ) is represented by the directed graph, $({\cal G}_0,{\cal G}_1)$, where ${\cal G}_0$,${\cal G}_1$ marks the set of vertices and directed edges of the graph (respectively).  If at some instance $t$, node $j$ has a queue of size $n$, one says that the
node is in the state represented by the ket-vector $|n\rangle$, where $n=0,1,\cdots$.
Then, any ``pure" state of the network will be denoted by the ket-vector
$|{\bm n}\rangle$, where the components $n_{i}$ of a vector ${\bm n}=(n_i|i\in {\cal G}_0)$ are labeled by the network nodes (vertices).
If a state $|{\bm n}\rangle$ is realized with the probability $P({\bm n})$, we
say that the entire network is in the following ``mixed" state
\begin{equation}
 |s\rangle =\sum_{\bm n} P({\bm n}) |{\bm n}\rangle, \quad
 \sum_{\bm n} P({\bm n})=1,
 \label{state}
 \end{equation}
where the last condition reflects the fact that the total probability equals unity.
Here and below we formally assume that $P({\bm n})=0$ whenever any component of the vector ${\bm n}$ is negative.

It is convenient to introduce a Hilbert space of ${\cal G}_0$-dimensional analytic functions of the vector variable ${\bm z}=(z_i|i\in{\cal G}_0)$
\begin{eqnarray}
{\cal P}({\bm z})=\sum_{\bm n} P({\bm n})\prod_{i\in{\cal G}_0} z_i^n,
\label{gen}
\end{eqnarray}
which is also known as the generating function in the theory of birth-death processes \cite{75Roz,83Gar}.

The ``quantum" (pure) states are transformed by the following creation and annihilation operators:
\begin{eqnarray}
\hat{a}^+_j|\cdots,n_j,\cdots\rangle=|\cdots,n_j+1,\cdots\rangle,\quad
\hat{a}_j|\cdots,n_j,\cdots\rangle=n_j|\cdots,n_j-1,\cdots\rangle.
\label{aa}
\end{eqnarray}

The normalization condition in Eq.~(\ref{state}), i.e. conservation of probability, reads
 \begin{equation}
 \langle {\bm 0} | \exp\left(\sum_{j\in{\cal G}_0} \hat{a}_j\right) |s\rangle =1,
 \label{norma}
 \end{equation}
where the vacuum state $|{\bm 0}\rangle\equiv |0,\cdots,0\rangle$  corresponds to the empty queue over the entire network.

In these notations ME becomes
 \begin{equation}
 \partial_t |s\rangle =\hat{H} |s\rangle \,,
 \label{doi}
 \end{equation}
where $\hat{H}$ is the Hamiltonian operator of the Q-network. In an integrated form, Eq.~(\ref{doi}) is equivalent to
\begin{eqnarray}
|s(t)\rangle=\hat{U}(t)|s(0)\rangle,\quad \hat{U}(t)\equiv T\exp\left(\int_0^tdt'\hat{ H}\right),
\label{doi_exp}
\end{eqnarray}
where $T\exp$ is defined as a time-ordered exponential, i.e. the product of time-discretized operators.  Furthermore, it becomes normal exponential if the parameters of $\hat{H}$ (i.e. transition rates) do not carry explicit time dependence.

Note that the Hamiltonian $\hat{H}$ is always real, as it represents probabilities which are positive and bounded.  Thus the normalization conditions
(\ref{norma}) and Eq.~(\ref{doi}), which should be enforced by the theory for any feasible ${\bm n}$, result in
 \begin{eqnarray}
 \langle {\bm 0} | \exp\left(\sum_j \hat{a}_j\right)\hat{ H}=0,
 \label{constr}
 \end{eqnarray}
where we have used standard bra-vector notations.  Recall that an operator acting on the bra-vector from the right generates a bra-vector, and all the features of left operations can be extracted directly from the normal definition of the bra-(c)-ket scalar product, $\forall n,m:\quad\langle n|m\rangle=\delta(n,m)$.
Stating it differently,  $\langle {\bm 0} | \exp\left(\sum_j \hat{a}_j\right)$ is the left eigen-vector of the Hamiltonian with zero eigen-value.

The expectation value of a (dummy) operator $\hat{\bullet}$ over a state $|s\rangle$ is
 \begin{eqnarray}
 \langle \hat{\bullet} \rangle \equiv
 \langle {\bm 0} | \exp\left(\sum_{j\in{\cal G}_0} \hat{a}_j\right) \hat{\bullet} |s\rangle,
 \label{aver}
 \end{eqnarray}
and according to Eq.~(\ref{constr}), the corresponding Heisenberg (evolution) equation becomes
 \begin{eqnarray}
 \partial_t \langle \hat{\bullet} \rangle
 =\langle [\hat{\bullet},\hat{H}] \rangle.
 \label{heis}
 \end{eqnarray}
Here we have assumed that $\hat{\bullet}$ does not have an explicit time dependence, and $[\hat{A},\hat{B}]$ is the standard notation for a commutator.

\subsection{$M/M/\infty$ network}
\label{subsec:MMinfty}

The generic form of the ME for a network of $M/M/\infty$ queues is
\begin{eqnarray}
&& \frac{\partial}{\partial t} P({\bm n};t)=\sum_{(i,j)\in{\cal G}_1}^{i,j\neq 0}\lambda_{ij}\left((n_i+1)P(\cdots,n_i+1,\cdots,n_j-1,\cdots;t)-
n_iP(\cdots,n_i,\cdots,n_j,\cdots;t)\right)\nonumber\\
&& +\sum_{i\in{\cal G}_0} \lambda_{0i}\left(P(\cdots,n_i-1,\cdots;t)-P(\cdots,n_i,\cdots;t)\right)\nonumber\\ &&
+\sum_{i\in{\cal G}_0}\lambda_{i0}\left((n_i+1)P(\cdots,n_i+1,\cdots;t)-n_iP(\cdots,n_i,\cdots;t)\right).
\label{ME}
\end{eqnarray}
Here $(i,j)$ stands for the directed edge of the network corresponding to a job transfer from site $i$ to site $j$, with Poisson rate $\lambda_{ij}$; and $\lambda_{0j}$,$\lambda_{j0}$ are the Poisson rates for job injection and removal to/from the network at site $j$ (respectively).  Applying summation over properly weighted ${\bm n}$-states to both sides of Eq.~(\ref{ME}), and using
the relations Eq.~(\ref{doi},\ref{doi_exp}), one arrives at the following Hamiltonian:
\begin{eqnarray}
\hat{H}_\infty=\sum_{(i,j)\in{\cal G}_1}^{i,j\neq 0} \lambda_{ij} (\hat{a}^+_j-\hat{a}^+_i)\hat{a}_i
+\sum_{i\in{\cal G}_0} \left(\lambda_{0i} (\hat{a}^+_i-1) +\lambda_{i0} (1-\hat{a}_i^+)\hat{a}_i\right),
\label{H}
\end{eqnarray}
which was first derived for the problem in \cite{85Pel}.

We further introduce a path-integral representation. The analytic structure of the theory is as follows:
\begin{eqnarray}
&& {\cal P}({\bm z};t)=\int\frac{d{\bm \zeta}d{\bm \zeta}'}{(2\pi i)^{|{\cal G}_0|}}W({\bm z},{\bm\zeta})
{\cal P}({\bm\zeta}';0)\exp(-{\bm \zeta}{\bm\zeta}'),\label{PI1}\\
&& W({\bm z},{\bm\zeta})=\int^{{\bm \eta}'(t)={\bm z}}_{{\bm \eta}(0)={\bm\zeta}}
{\cal D}{\bm\eta}{\cal D}{\bm\eta}'
\exp\left({\bm z}{\bm\eta}(t)-\int_0^t dt'\left({\bm\eta}'(t')\dot{\bm\eta}(t')-{\cal H}_\infty({\bm\eta}'(t'),{\bm\eta}(t))\right)\right),\label{PI2}
\end{eqnarray}
where ${\cal H}_\infty({\bm \eta}',{\bm\eta})$ corresponds to $\hat{ H}_\infty$ expressed as a polynomial over the creation/annihilation operators, so that the creation operators are all positioned on the left from the annihilation operators (normal ordering), and $\hat{a}^+_j$,$\hat{a}_j$ are replaced by $\eta_j'$ and $\eta_j$ (respectively).  Thus, for a general birth-death model (\ref{H}) over a network ${\cal G}$, one derives
\begin{eqnarray}
{\cal H}_\infty({\bm\eta}',{\bm\eta})=\sum_{(i,j)\in{\cal G}_1}^{i,j\neq 0} \lambda_{ij} (\eta_j'-\eta_i')\eta_i
+\sum_{i\in{\cal G}_0} \left(\lambda_{0i} (\eta_i'-1) +\lambda_{i0} (1-\eta_i')\eta_i\right).
\label{Heta}
\end{eqnarray}
As usual, the path-integrals in Eq.~(\ref{PI2}) should be understood as the continuous limit of the following discretized multiple integral (see \cite{85Pel} for explanations and accurate
validation of the proper discrete-time regularization):
\begin{eqnarray}
\label{PI2_discr}
&& W({\bm z},{\bm\zeta})=\lim_{N\to\infty}
\int \prod_{l=1}^{N-1}\frac{d{\bm \eta}_l d{\bm\eta}_l'}{(2\pi i)^{|{\cal G}_0|}}\\
&& \times\exp\left({\bm z}{\bm\eta}_{N-1}+\Delta {\cal H}({\bm z},{\bm\eta}_{N-1})+\sum_{l=1}^{N-1}\left(-{\bm\eta}_l'({\bm\eta}_l-{\bm\eta}_{l-1})+\Delta {\cal H}({\bm \eta}'_l,{\bm\eta}_{l-1})\right)\right),\nonumber
\end{eqnarray}
where $\Delta=t/N$.

Finally,  one finds that the creation-annihilation (\ref{H}) and the path-integral (\ref{PI2},\ref{PI2_discr}) formulations of the birth-death process also allow for the following simple ``differential" interpretation for the analytic function defined in (\ref{gen}):
\begin{eqnarray}
&& \partial_t{\cal P}({\bm z};t)=\hat{\cal H}_\infty({\bm z}){\cal P}({\bm z};t),
\label{PzHz}\\
&& \hat{\cal H}_\infty({\bm z})={\cal H}_\infty({\bm z},\partial_{\bm z})=\sum_{(i,j)\in{\cal G}_1}^{i,j\neq 0} \lambda_{ij} (z_j-z_i)\partial_{z_i}
+\sum_{i\in{\cal G}_0} \left(\lambda_{0i} (z_i-1) +\lambda_{i0} (1-z_i)\partial_{z_i}\right).
\label{Hz}
\end{eqnarray}
Thus the mapping from the creation-annihilation operators to the poly-differential operators (holomorphic representation) is
\begin{eqnarray}
\left(\hat{a}^+,\hat{a}\right)\to \left(z,\partial_z\right).
\label{a-to-z}
\end{eqnarray}

Looking for a steady state (time-independent) solution of Eq.~(\ref{Hz}) in the exponential form
\begin{eqnarray}
{\cal P}_\infty({\bm z})=\exp\left[\sum_{i\in{\cal G}_0} h_i (z_i-1)\right],
\label{exp}
\end{eqnarray}
and substituting the ansatz into Eq.~(\ref{PzHz}), one arrives at the following set of conditions on ${\bm h}$:
\begin{eqnarray}
&& \sum_{i\in {\cal G}_0}\left(\lambda_{0i}-\lambda_{i0} h_i\right)=0,\label{global}\\
&& \forall i\in{\cal G}_0:-\quad h_i\sum_{j\neq 0}^{(i,j)\in{\cal G}_1}\lambda_{ij}+\sum_{j\neq 0}^{(j,i)\in{\cal G}_1}\lambda_{ji}h_j+\lambda_{0i}-\lambda_{i0} h_i=0.\label{local_i}
\end{eqnarray}
Although it seems that there is one more condition than the number of variables, the conditions are dependent (sum Eq.~(\ref{local_i}) over all vertices of the graph).  Therefore, solving the system of inhomogeneous linear equations Eq.~(\ref{local_i}), which we restate for convenience as
\begin{eqnarray}
\hat{\Lambda}{\bm h}={\bm \lambda}_{in},\quad {\bm \lambda}_{in}\equiv(-\lambda_{0i}|i\in{\cal G}_0),\quad
\hat{\Lambda}=(\Lambda_{ij}|i,j\in{\cal G}_0),\quad\Lambda_{ij}=\left\{\begin{array}{cc}
                      -\lambda_{i0}-\sum_k^{(i,k)\in{\cal G}_1}\lambda_{ik}, & i=j\\
                      \lambda_{ji},& i\neq j,\end{array}\right.,
\label{Lin_Eq}
\end{eqnarray}
consists in evaluating
\begin{eqnarray}
{\bm h}=\hat{\Lambda}^{-1}{\bm\lambda}_{in}.
\label{h_Lambda_Inv}
\end{eqnarray}
Here the existence of the steady state solution requires that: (a) $\hat{\Lambda}$ is not singular, and (b) all components of the ${\bm h}$ vector that solves Eq.~(\ref{h_Lambda_Inv}) are positive.

Note that the form of Eq.~(\ref{exp}) is fully factorized.  Thus once the valid solution of Eq.~(\ref{local_i}) is found, the full probability of observing the system in any given state is decomposed into a product of probabilities, each evaluated at the relevant graph vertex. Recall that this occurs in spite of the fact that to find the re-normalized rates $h_i$ one must solve a graph-global linear problem. This strong symmetry of the Poisson-In-Poisson-Out process, observed in spite of the fact that the DB is broken, is referred to (in the Queuing Theory literature) as ``quasi"-DB \cite{79Kel,01CY}.

The special feature (memoryless property) of the exponential distribution is also very transparent in the creation-annihilation language. Indeed, one observes that the exponential (in quantum mechanics also referred to as ``coherent" ) state $\exp(h\hat{a}^+)|0\rangle$ is the eigen-function of the annihilation operator $\hat{a}$, with the eigen-value $h$
\begin{eqnarray}
\hat{a}|\mbox{coh}_\infty(h)\rangle=h|\mbox{coh}_\infty(h)\rangle,\quad |\mbox{coh}_\infty(h)\rangle\equiv\exp(h\hat{a}^+)|0\rangle.
\label{coh}
\end{eqnarray}
Therefore,
\begin{eqnarray}
\hat{H}_\infty|\mbox{coh}_\infty({\bm h})\rangle =\left(\sum_{(i,j)\in{\cal G}_1}^{i,j\neq 0} \lambda_{ij} (\hat{a}^+_j-\hat{a}^+_i)h_i
+\sum_{i\in{\cal G}_0} \left(\lambda_{0i} (\hat{a}^+_i-1) +\lambda_{i0} (1-\hat{a}_i^+)h_i\right)\right)|\mbox{coh}_\infty({\bm h})\rangle,
\label{H_coh}
\end{eqnarray}
and the stationarity condition $\hat{H}_\infty|\mbox{coh}_\infty({\bm h})\rangle=0$ translates exactly into Eqs.~(\ref{local_i}), where the $i$-th equation correspond to the condition that the $c$-factor in front of the corresponding $\hat{a}^+_i$ is zero.

\subsection{$M/M/1$ network}
\label{subsec:MM1}

Consider a network of $M/M/1$ processes. In this case the ME adopts the following form:
\begin{eqnarray}
&& \frac{\partial}{\partial t} P({\bm n};t)=\sum_{(i,j)\in{\cal G}_1}^{i,j\neq 0}\lambda_{ij}\Biggl(P(\cdots,n_i+1,\cdots,n_j-1,\cdots;t)-P(\cdots,n_i,\cdots,n_j,\cdots;t)\Biggr)\nonumber\\ && +\sum_{i\in{\cal G}_0}\lambda_{0i}\left(P(\cdots,n_i-1,\cdots;t)-P(\cdots,n_i,\cdots;t)\right)\nonumber\\ &&
+\sum_{i\in{\cal G}_0}\lambda_{i0}\left(P(\cdots,n_i+1,\cdots;t)-P(\cdots,n_i,\cdots;t)\right).
\label{ME1}
\end{eqnarray}
Here $\theta(x)$ is the characteristic function of the logical condition $x$, i.e. it is unity when the condition is satisfied and zero otherwise.
The corresponding Hamiltonian operator in Eq.~(\ref{doi},\ref{doi_exp}) is of the form
\begin{eqnarray}
\hat{H}_1=\sum_{(i,j)\in{\cal G}_1}^{i,j\neq 0} \lambda_{ij} (\hat{a}^+_j-\hat{a}^+_i)\hat{b}_i
+\sum_{i\in{\cal G}_0} \left(\lambda_{0i} (\hat{a}^+_i-1) +\lambda_{i0} (1-\hat{a}_i^+)\hat{b}_i\right).
\label{H1}
\end{eqnarray}
Here $\hat{b}_i$ is a ``skewed" annihilation operator (see e.g. \cite{07Zei} for a similar operational rule), such that
$\hat{b}_i|n_i\rangle=\theta(n_i>0)|n_i-1\rangle$, and
in both Eq.~(\ref{ME1}) and Eq.~(\ref{H1}) we keep the same notation as in Eqs.~(\ref{ME},\ref{H}) (respectively).

Note that $\hat{b}$ is expressed in terms of $\hat{a}$ and $\hat{a}^+$ in an extremely nonlinear way. However, the representation allows a simple ``analytic" interpretation \cite{07Zei}:
\begin{eqnarray}
\hat{b}\sum_n p_n|n\rangle \to \frac{p(z)-p(0)}{z},\quad \mbox{where}\quad  p(z)=\sum_n p_n z^n.
\label{b-to-z}
\end{eqnarray}
For the introduced generating function representation, the analog of Eqs.~(\ref{PzHz},\ref{Hz}) becomes
\begin{eqnarray}
&& \partial_t{\cal P}({\bm z};t)=\sum_{(i,j)\in{\cal G}_1}^{i,j\neq 0} \lambda_{ij} (z_j-z_i)
\frac{{\cal P}({\bm z};t)-{\cal P}({\bm z}_{\sim i};t)}{z_i}
+\sum_{i\in{\cal G}_0} \lambda_{0i} (z_i-1){\cal P}({\bm z};t)\nonumber\\
&& +\sum_{i\in{\cal G}_0}\lambda_{i0} (1-z_i)
\frac{{\cal P}({\bm z};t)-{\cal P}({\bm z}_{\sim i};t)}{z_i}.
\label{Pz1}
\end{eqnarray}
Here ${\bm z}_{\sim i}\equiv ((1-\delta_{ij})z_j|j\in{\cal G}_0)$.  In words, this is the vector ${\bm z}$ with the component $z_i$ replaced by zero.

Let us come back to the skewed-creation-annihilation representation, and note that the coherent states associated with this ``skewed" annihilation operator $\hat{b}$ were discussed in \cite{07Zei}.  The approach can also be traced back in the Queuing Theory literature to the classic papers of Massey \cite{84Mas_a,84Mas_b} on the operator approach to Jackson networks.
The coherent states for $\hat{b}$ are constructed as follows:
\begin{eqnarray}
\hat{b}|\mbox{coh}_1(h)\rangle=h|\mbox{coh}_1(h)\rangle,\quad |\mbox{coh}_1(h)\rangle\equiv\frac{1}{1-h \hat{a}^+}|0\rangle.
\label{coh1}
\end{eqnarray}
Then the analog of Eq.~(\ref{H_coh}) becomes
\begin{eqnarray}
\hat{H}_1|\mbox{coh}_1({\bm h})\rangle =\left(\sum_{(i,j)\in{\cal G}_1} \lambda_{ij} (\hat{a}^+_j-\hat{a}^+_i)h_i
+\sum_{i\in{\cal G}_0} \left(\lambda_{0i} (\hat{a}^+_i-1) +\lambda_{i0} (1-\hat{a}_i^+)h_i\right)\right)|\mbox{coh}_1({\bm h})\rangle.
\label{H_coh1}
\end{eqnarray}
Furthermore, the condition of stationarity, $\hat{H}_1|\mbox{coh}_1({\bm h})\rangle=0$, translates exactly into Eqs.~(\ref{local_i}), where the $i$-th equation corresponds to the condition that the $c$-factor in front of the respective $\hat{a}^+_i$ is zero.
We conclude that the stationary distribution of the $M/M/1$-network is
\begin{eqnarray}
{\cal P}_1({\bm z})=\prod_{i\in{\cal G}_0}\frac{1-h_i}{1-h_i z_i},
\label{stat1}
\end{eqnarray}
where ${\bm h}$ is the solution of Eqs.~(\ref{local_i}).

\subsection{$M/M/m$ network}
\label{subsec:MMm}

The ME in the general case of an inhomogeneous $M/M/m$-network, with positive integers $m_i$ (number of tellers) assigned to each vertex $i$ of the graph, can be represented by
\begin{eqnarray}
&& \frac{\partial}{\partial t} P({\bm n};t)=\sum_{(i,j)\in{\cal G}_1}\lambda_{ij}\Biggl(\theta_{m_i}(n_i+1)\theta(n_j>0)P(\cdots,n_i+1,\cdots,n_j-1,\cdots;t)\nonumber\\
&&-\theta_{m_i}(n_i)
P(\cdots,n_i,\cdots,n_j,\cdots;t)\Biggr)\nonumber\\
&& +\sum_{i\in{\cal G}_0} \lambda_{0i}\left(\theta(n_i>0)P(\cdots,n_i-1,\cdots;t)-P(\cdots,n_i,\cdots;t)\right)\nonumber\\ &&
+\sum_{i\in{\cal G}_0}\lambda_{i0}\left(\theta_{m_i}(n_i+1)P(\cdots,n_i+1,\cdots;t)-\theta_{m_i}(n_i)P(\cdots,n_i,\cdots;t)\right),
\label{MEm}\\
&& \theta_m(n)=\min(n,m).
\label{theta_m_n}
\end{eqnarray}
The evolution operator (Hamiltonian) becomes
\begin{eqnarray}
&& \hat{H}=\sum_{(i,j)\in{\cal G}_1}^{i,j\neq 0} \lambda_{ij} (\hat{a}^+_j-\hat{a}^+_i)\hat{b}_i^{(m_i)}
+\sum_{i\in{\cal G}_0} \left(\lambda_{0i} (\hat{a}^+_i-1) +\lambda_{i0} (1-\hat{a}_i^+)\hat{b}_i^{(m_i)}\right),
\label{H_gen}\\
&& \hat{b}^{(m)}|n\rangle=\theta_m(n)|n-1\rangle.
\label{bmn}
\end{eqnarray}
Thus the problem of finding the stationary solution is reduced (pretty much like before in the $m=\infty$ and $m=1$ cases) to constructing coherent states for the annihilation operator $\hat{b}^{(m)}$:
\begin{eqnarray}
&& \hat{b}^{(m)}|\mbox{coh}_m(h)\rangle=h|\mbox{coh}_m(h)\rangle,\quad |\mbox{coh}_m(h)\rangle\equiv
g_m(h\hat{a}^+)|0\rangle,\label{cohm1}\\
&&
g_m(x)\equiv\sum_{k=0}^\infty \frac{x^k}{\prod_{l=1}^k\theta_m(l)}=
\frac{m^m}{m!}\frac{1}{1-x/m}+\sum_{k=0}^{m-1}x^k\left(\frac{1}{k!}-\frac{m^{m-k}}{m!}\right).
\label{cohm2}
\end{eqnarray}
Finally, the full expression for the generating function of the stationary solution over the general network becomes
\begin{eqnarray}
{\cal P}({\bm z})=\prod_{i\in{\cal G}_0}\frac{g_{m_i}(h_i z_i)}{g_{m_i}(h_i)},
\label{statG}
\end{eqnarray}
where ${\bm h}$ is the solution of Eqs.~(\ref{local_i}).
Therefore, by Eq.~(\ref{gen}),
\begin{eqnarray}
P({\bm n})=Z^{-1}\prod_{i\in{\cal G}_0}\frac{h_i^{n_i}}{\prod_{l_i=1}^{n_i}\theta_{m_i}(l_i)}.
\label{Pm_gen}
\end{eqnarray}
Obviously, Eqs.~(\ref{statG},\ref{Pm_gen}) are consistent with Eq.~(\ref{exp}) and Eq.~(\ref{stat1}) when ${\bm m}=(m_i|i\in{\cal G}_0)$ is set to ${\bm m}={\bm \infty}$ and ${\bm m}={\bm 1}$ (respectively).

Note (for the sake of accurateness) that when deriving Eq.~(\ref{Pm_gen}) in the operator formalism we took advantage of the important fact that both left (bra-) and right (ket-) zero eigenvalues of the Hamiltonian (\ref{H_gen}), described by
$\langle {\bm 0} | \exp\left(\sum_j \hat{a}_j\right)$ and $\prod_{i\in{\cal G}_0}g_{m_i}(h_i\hat{a}^+)|{\bm 0}\rangle$ respectively, are explicitly known.

Note that one can recalculate any moment of $n$ from either Eq.~(\ref{statG}) or Eq.~(\ref{Pm_gen}).  In particular, for the first moment at a station we arrive at
\begin{equation}
\langle n_i\rangle=\frac{
\langle {\bm 0} | \exp\left(\sum_{j\in{\cal G}_0} \hat{a}_j\right) \hat{a}^+_i\hat{a}_i\prod_{k\in{\cal G}_0} g_{m_k}(h_k\hat{a}^+)|{\bm 0}\rangle}
{\langle {\bm 0} | \exp\left(\sum_{j\in{\cal G}_0} \hat{a}_j\right)\prod_{k\in{\cal G}_0} g_{m_k}(h_k\hat{a}^+)|{\bm 0}\rangle}
=
\left.\frac{\partial}{\partial z_i}\frac{g_{m_i}(h_iz_i)}{g_{m_i}(h_i)}\right|_{{\bm z}=1},
\label{n=1}
\end{equation}
where $g_m(x)$ is taken from Eq.~(\ref{cohm2}) and (as before) ${\bm h}$ is the solution of Eqs.~(\ref{local_i}). Note that the moments are finite only if $h_i<m_i$, which thus defines the condition for Q-network stability (statistical stationarity) \cite{01CY}.

\section{Statistics of Network Currents}
\label{sec:Currents}

A (general) queueing network is naturally characterized by the actual current of particles/jobs going through and being processed at each station according to a certain service discipline $\big($ e.g. First-In-First-Out (FIFO) $\big)$.  Here the quasi-current characterizes the activities of tellers, not focusing on the dynamics of the individual particles at all. In other words, actual current tracks the dynamics of the jobs/particles, while quasi-current tracks quasi-particles/jobs assuming that all the jobs waiting for service at a station are fully equivalent and not prioritized.
Actual current and quasi-current coincide in the case of an $M/M/\infty$ network, when the individual jobs do not interact at all, as well as any network in which all particles (customers) are identical. With this disclaimer, we will be discussing quasi-currents for the remainder of the paper, and refer to them as currents to simplify our exposition.

As shown below, calculating the statistics of the currents in a Jackson network becomes tractable in the two regimes that we will study: the ``uncongested'' and ``congested'' regimes.

In the ``uncongested'' regime, this arises from the product-form symmetry characterizing the regime (an extension of the product-form symmetry discussed earlier), as well as the ${\bm m}$-independence property (universality).  Note that for networks without feedback, the statement of ${\bm m}$-independence is equivalent to the fact that in steady state, the flow along arcs will be Poisson (with rate independent of the number of servers).  As mentioned previously, this phenomenon was discovered earlier in the Queueing Theory literature \cite{81WV,90MW,77BM}.  However (and to the best of our knowledge), the extension of this statement to the asymptotic (large time) limit for more general networks (unscaled, not in the fluid limit or hydrodynamic limit) has not been formally explored in the literature.

The remainder of this Section is partitioned into four Subsections. We start from a general discussion of the current related objects in Section \ref{subsec:gen_currents}. In Section \ref{subsec:uncongested} we consider the uncongested regime.  We also develop several generalizations of the Doi-Peliti technique to account for currents, and develop some machinery necessary for the statement of our results.  At the end of the Section,
our analysis naturally leads to the identification of the uncongested regime's breakdown, namely the identification of a phase transition in the space of currents.  We also show that this transition is second-order, thus translating into smoothness of the Cr\'{a}mer function of currents (continuity of the first and the second derivatives) at the transition.  In Section \ref{subsec:congested}, we extend the coherent state formalism to the ``congested'' regime via a simple reduction of the network graph.  We note that a similar decomposition was applied in \cite{Phys2RH08}, and is similar in spirit to many such reductions appearing throughout the Queueing Theory literature \cite{M99,AFM09,DNR94,ES05,ES04,VCW94,BPT98}.

Finally, in Section \ref{subsec:feedback} we illustrate the general theory using our enabling example of a single node with feedback.

\subsection{Preliminary General Remarks}
\label{subsec:gen_currents}

We will mainly be interested to evaluate the joint distribution function of the currents and queue sizes where the latter are averaged over the entire time horizon. In the following we will use $P(\bar{{\bm n}},{\bm J}|t)$ notation for the main object of interest.  However it is technically more convenient to start from another (and to a degree auxiliary) object defined as a joint distribution function of currents and queues where the latter are observed at the final moment of time.  We will see below that it is important to differentiate these generally distinct objects.

Let $P({\bm n}(t),{\bm J}|t)$ denote the joint probability distribution function of the queue size at the final moment of time $t$, ${\bm n}(t)=(n_i(t)|i\in{\cal G}_0)$, and currents accumulated over the $[0;t]$ interval of time, ${\bm J}=(J_{ij}|(i,j)\in{\cal G}_1)$, where the latter are defined on all edges of the graph and the former are defined (as before) on vertices.  The ME for this object is the natural generalization of Eq.~(\ref{MEm}), which we now present in operator form (to allow for more compact notations).  In particular, the operator form of the ME for $P({\bm n}(t),{\bm J}|t)$ is
\begin{eqnarray}
&& \partial_t |{\bm s}({\bm n}(t);{\bm J})\rangle=
\left(\hat{ H}+\sum_{(i,j)\in{\cal G}_1}\hat{\it J}_{ij}\right)|{\bf s}({\bm n}(t);{\bm J})\rangle,
\label{H-current}\\
&& \forall i,j\neq 0:\quad\hat{\it J}_{ij}=\lambda_{ij}(\hat{a}_{ij}^+-1)\hat{a}^+_j\hat{b}_i^{(m_i)},
\label{current}\\
&& \hat{\it J}_{0i}=\lambda_{0i}(1-\hat{a}_{0i}^+)\hat{a}_i^+,\quad
\hat{\it J}_{i0}=\lambda_{i0}(\hat{a}_{i0}^+-1)\hat{b}_i^{(m_i)},
\label{current0}
\end{eqnarray}
where $\hat{H}$ is defined in Eq.~(\ref{H_gen}).  Here we have assumed that the currents are discrete and positive, and the respective ket-vector is related to the joint PDF of ${\bf n}(t)$ and ${\bf J}$ as follows:
\begin{eqnarray}
|{\bm s}({\bm n}(t);{\bm J})\rangle=P({\bm n}(t);{\bm J})|{\bm n};{\bm J}\rangle.
\label{s-ket}
\end{eqnarray}
$\hat{\it J}_{ij}$ $\big($ in\ Eq.~(\ref{current}) $\big)$ is the operator for the amount of current from site $i$ to site $j$, and $\hat{a}_{ij}^+$ is the newly introduced creation operator $\big($ at edge $(i,j)$ $\big)$ acting on the space of discrete positive currents.  We define operators for incoming and outgoing currents in Eq.~(\ref{current0}) similarly.
Formal solution of Eq.~(\ref{H-current}) is
\begin{eqnarray}
|{\bf s}({\bm n}(t);{\bm J})\rangle=\exp\left(t
\left(\hat{ H}+\sum_{(i,j)\in{\cal G}_1}\hat{\it J}_{ij}\right)\right)|{\bf s}({\bm n}(0);{\bm J})\rangle, \label{formal_s_n}
\end{eqnarray}
where the ket-state on the rate correspond to the ``initial" steady distribution of queues, described by Eq.~(\ref{Pm_gen}), and zero initial current:
$|{\bf s}({\bm n}(0);{\bm J})\rangle=|s\rangle\otimes|{\bm J}={\bm 0}\rangle$, where we follow notations introduced in the introduction and $|s\rangle=\sum_{\bm n}P({\bm n})|{\bm n}\rangle$.

It follows that the generating function over the currents $(i,j)\in{\cal G}_1$, accounting for
the incoming $\big( (0,i)\in{\cal G}_1 \big)$ and outgoing $\big( (i,0)\in{\cal G}_1 \big)$ arcs, is
\begin{eqnarray}
|{\bm s}_{\bm q}({\bm n}(t))\rangle=\sum_{\bm J}
\prod_{(i,j)\in {\cal G}_1}q_{ij}^{J_{ij}}|{\bm s}({\bm n}(t);{\bm J})\rangle.
\label{curr-gen}
\end{eqnarray}
According to our standard birth-death [creation/annihilation] rules, the object described by Eq.~(\ref{curr-gen}) satisfies
\begin{eqnarray}
&& \partial_t |{\bm s}_{\bm q}({\bm n}(t))\rangle=\hat{H}_{\bm q}
|{\bm s}_{\bm q}({\bm n}(t))\rangle,\quad |{\bm s}_{\bm q}({\bm n}(t))\rangle=\exp(t\hat{H}_q)\sum_{\bm n}P({\bm n})|{\bm n}\rangle=\exp(t\hat{H}_q)|s\rangle,
\label{H-current}\\
&& \hat{H}_{\bm q}=\sum_{(i,j)\in{\cal G}_1} \lambda_{ij} (\hat{a}^+_j-\hat{a}^+_i)\hat{b}_i^{(m_i)}
\nonumber\\ &&
+\sum_{(i,j)\in{\cal G}_1}^{i\neq 0,j\neq 0}\lambda_{ij}(q_{ij}-1)\hat{a}^+_j\hat{b}_i^{(m_i)}+
\sum_{(0,i)\in{\cal G}_1}\lambda_{0i}(q_{0i}a_i^+-1)+
\sum_{(i,0)\in{\cal G}_1}\lambda_{i0}(q_{i0}-a_i^+)\hat{b}_i^{(m_i)}.\label{Hq}
\end{eqnarray}

Note that even though our evaluation in Eq.~(\ref{H-current}) applies to the general object, namely to the joint distribution function of currents over the entire network, one may compute the relevant marginals (say the distribution function for the current over a single edge) in a straightforward manner.  We also note that the operator $\hat{H}_{\bm q}$ can be obtained by modifying/twisting the evolution operator $\hat{H}$, given by Eq.~(\ref{bmn}), as follows.  One weights the off-diagonal terms of $\hat{H}$ in the space of populations $|{\bm n}\rangle$, namely $\lambda_{ij}\hat{a}^+_j\hat{b}_i^{(m_i)}$, $\lambda_{0i}a_{i}^{+}$, and $\lambda_{i0}\hat{b}_{i}^{(m_{i})}$, with the factors $q_{ij}$, $q_{0i}$, and $q_{i0}$ respectively.  This corresponds to viewing the underlying stochastic process as a Markov chain on an infinite graph, whose nodes are labeled by pure states $|{\bm n}\rangle$, whereas links represent the set of processes allowed by the evolution operator $\hat{H}$. Within such a picture, the set of ${\bm q}$ parameters plays the role of discrete gauge fields (vector potentials), and $\hat{H}_{{\bm q}}$ is interpreted as the evolution operator, ``twisted'' by the gauge field ${\bm q}$, as described in \cite{07CCMT}. Since here we are dealing with oriented graphs, no constraints are imposed on ${\bm q}$.

It is also useful to consider the distribution function of queue at the finite moment of time, conditioned to specific value of the current generating parameter ${\bm q}$. This object, and the respective first moment of queue size, become
\begin{eqnarray}
&& P_{\bm q}({\bm n}(t))\propto \langle {\bm n}(t)|\exp\left(t\hat{H}_q\right)|s\rangle,
\label{Pqn(t)}\\
&& \langle n_i(t)\rangle_{\bm q}=\frac{\langle {\bm 0}|\exp\left(\sum_{j\in{\cal G}_0}\hat{a}_j\right)\hat{a}^+_i\hat{a}_i\exp\left(t\hat{H}_q\right)|s\rangle}{\langle {\bm 0}|\exp\left(\sum_{j\in{\cal G}_0}\hat{a}_j\right)\exp\left(t\hat{H}_q\right)|s\rangle}.
\label{n(t)}
\end{eqnarray}

Returning back to our main object of interest (the joint distribution function of the current and of the queue averaged over the time horizon) and following the same formalism/notations,  we derive the analogs of Eqs.~(\ref{Pqn(t)},\ref{n(t)})
\begin{eqnarray}
&&
P_{\bm q}(\bar{\bm n})\propto t^{-1}
\int_0^t dt'\langle {\bm 0}|\exp\left(\sum_{j\in{\cal G}_0}\hat{a}_j\right)\exp\left((t-t')\hat{H}_q\right)|{\bm n}(t')\rangle\langle {\bm n}(t')|
\exp\left(t'\hat{H}_q\right)|s\rangle,
\label{Pqbar_n}\\
&& \langle \bar{n}_i\rangle_{\bm q}= \frac{\int_0^t dt'\langle {\bm 0}|\exp\left(\sum_{j\in{\cal G}_0}\hat{a}_j\right)\exp\left((t-t')\hat{H}_q\right)\hat{a}^+_i\hat{a}_i
\exp\left(t'\hat{H}_q\right)|s\rangle}{\int_0^t dt'\langle {\bm 0}|\exp\left(\sum_{j\in{\cal G}_0}\hat{a}_j\right)\exp\left(t\hat{H}_q\right)|s\rangle}.
\label{bar_n}
\end{eqnarray}

\subsection{Uncongested Regime}
\label{subsec:uncongested}

We are now in a position to introduce the ``uncongested" regime on a formal level.  We will characterize this regime in terms of the existence of a special ``universal" product-form solution to Eq.~(\ref{H-current}), as well as the finiteness of a particular expectation value (capturing the fact that no condensation of particles occurs at any nodes), at sufficiently large observational time $t$.  Note that in the spirit of our operator-theoretic framework, we define the regime in terms of \textbf{both} the queueing network \textbf{and} the vector ${\bm q}$ on which one evaluates the generating function for occupation numbers in the network.
Thus a given queueing network may be in the regime for some evaluations of its generating function (certain values of ${\bm q}$) but not for others.  This will then be related to belonging (or not belonging) to the regime for different types of large deviations (of current) through the standard Legandre transform, which maps the generating function (evaluated at different ${\bm q}$) to the  Cr\'{a}mer function (evaluated at different-sized deviations).

We say that a given queueing network is in the ``uncongested" regime for a given vector ${\bm q}$ if
the ket vector $|{\bm s}_{\bm q}({\bm n})$ is dominated by the ground state of $\hat{H}_q$, i.e. at sufficiently large time, $|{\bm s}_{\bm q}({\bm n})\rangle \sim\exp(-\Delta({\bm q})t)|\mbox{coh}_{\bm m}({\bm h}({\bm q}))$, holds. In other words, the spectrum of $\hat{H}_q$ is such that its ground state is separated from the excited states by a finite gap, and thus at the times much large than inverse value of the gap the solution is completely described by the ground state only,  thus providing respective universality.  We also require that $\langle \bar{n}_i\rangle_{\bm q} < \infty$ for all nodes $i$ of the network in the ``uncongested" regime, ensuring that the expected number of particles (size of the queue averaged over time) does not diverge at any nodes as $t \rightarrow \infty$.

The statistics of queueing networks in the ``uncongested" regime may be analyzed using the techniques from Section \ref{sec:DPM}.  In particular, we substitute $|{\bm s}_{\bm q}({\bm n})\rangle$ by $\sim\exp(-\Delta({\bm q})t)|\mbox{coh}_{\bm m}({\bm h}({\bm q}))$ in Eq.~(\ref{H-current}) to arrive at
\begin{eqnarray}
&& \sum_{i\in{\cal G}_0}\left(q_{0i}\lambda_{0i}-\lambda_{i0} q_{i0} h_i({\bm q})\right)=\Delta({\bm q}),\label{global_q}\\
&&
\forall i\in{\cal G}_0:\quad -h_i({\bm q})\sum_{j\in{\cal G}_0}^{(i,j)\in{\cal G}_1}\lambda_{ij}+\sum_{j\in{\cal G}_0}^{(j,i)\in{\cal G}_1}q_{ji}\lambda_{ji}h_j({\bm q})+\lambda_{0i}-\lambda_{i0} h_i({\bm q})=0.\label{local_i_q}
\end{eqnarray}
This set of relations generalizes the stationary (${\bm q}={\bm 1}$ ) relations (\ref{global},\ref{local_i}), and are thus consistent with $\Delta({\bm 1})=0$.  Eqs. ~(\ref{global_q},\ref{local_i_q}) describe the right (ket) eigen-function of the ground-state of the evolution operator/Hamiltonian (\ref{Hq}), while the corresponding left (bra)-eigenfunction is described in Appendix \ref{app:Left}.

Note that, replacing all internal $q$-variables by unity, the basic set of equations for ${\bm h}$ does not depend on the remaining $q_{0i}$ and $q_{i0}$ components.  It follows that the respective ${\bm h}$ are identical to the one derived before for the stationary ${\bm h}({\bm 0})$ (no currents) setting.
Moreover, $\Delta({\bm q}_0)=\sum_i\left((q_{0i}-1)\lambda_{0i}-\lambda_{i0} (q_{i0}-1) h_i({\bm 0})\right)$.  This observation translates (after the obvious Legandre transform) into the statement that all the currents entering and leaving the network are asymptotically Poisson, which (as already mentioned) was known previously in the Queuing literature \cite{77BM,79Kel}.

For the sake of simplicity, hereafter we exclude the incoming and outgoing currents from consideration, which is achieved by setting $q_{i0}=q_{0i}=1$.
The consistency of Eqs.~(\ref{global_q}) and Eqs.~(\ref{local_i_q}) (the two are just generalized versions of Eqs.~(\ref{local_i},\ref{global})) translates into the following expression for the lowest eigenvalue:
\begin{eqnarray}
\Delta({\bm q})=-\sum_{(i,j)\in{\cal G}_1}^{i,j\in{\cal G}_0}h_i({\bm q})\lambda_{ij}(q_{ij}-1).\label{Delta}
\end{eqnarray}

For sufficiently large $t$ (it is our Large Deviation Parameter and thus should at least be significantly larger than the correlation time of the system) we obtain the following asymptotic expression for $P_{\bm q}(t)=\sum_{\bm n}P_{\bm q}(\bar{\bm n})$,
\begin{eqnarray}
&& P_{\bm q}(t)=\Psi({\bm q})
\exp(-t\Delta({\bm q}))\sim\exp\left(t\sum_{(i,j)\in{\cal G}_1}^{i,j\in{\cal G}_0}
h_i({\bm q})\lambda_{ij}(q_{ij}-1)\right),
\label{G_ass}\\
&& \Psi({\bm q})\equiv\langle {\bm 0}|\exp\left(\sum_{i\in{\cal G}_0} \bar{h}_i({\bm q})\hat{a}_i\right)|\mbox{coh}_{\bm m}({\bm h}({\bm q})\rangle. \label{Psi_q}
\end{eqnarray}
Here, in evaluating the pre-exponential factor  $\Psi({\bm q})$, we have used the assumption that the main contribution into Eq.~(\ref{Pqbar_n})  originates from $t', t-t'\gg 1/\Delta({\bm q})$.
Thus, $\langle {\bm 0}|\exp(\sum_i \bar{h}_i({\bm q})\hat{a}_i)$ in Eq.~(\ref{Psi_q}) is the bra-vector defined as the left eigenvector of $\hat{H}_{{\bm q}}$ with the same eigenvalue $\Delta({\bm q})$ (see Appendix \ref{app:Left} for more details) \footnote{Note that expression for the analog of $\Psi({\bm q})$ correspondent to $P_{\bm q}({\bm n}(t))$ is significantly different: $\langle {\bm 0}|\exp\left(\sum_{i\in{\cal G}_0}\hat{a}_i\right)|\mbox{coh}_{\bm m}({\bm h}({\bm q})\rangle$.}.  Here certain time-independent pre-factors (which do not impact the asymptotics up to exponential order) are ignored on the rhs.
As we are interested in the statistics of the currents that scale (grow) linearly with $t$, one can replace the sum in Eq.~(\ref{curr-gen}) by an integral, invert the relation, and arrive at the following saddle-point (large-deviation) expression:
\begin{eqnarray}
&& P({\bm J}|t)\sim\int {\cal P}_{\bm q}(t)\prod_{(i,j)\in{\cal G}_1}^{i,j\in{\cal G}_0}q_{ij}^{-J_{ij}}
\sim \exp\left(-t {\cal S}({\bm J}/t)\right),\nonumber\\
&&
{\cal S}({\bm  j})=
\sum_{(i,j)\in{\cal G}_1}^{i,j\in{\cal G}_0} \left(j_{ij}\ln(q_{ij}^*)+h_i({\bm q}^*)\lambda_{ij}(1-q_{ij})\right),
\label{P_J1}\\
&& \forall (k,l)\in{\cal G}_1\ \& k,l\in{\cal G}_0:\quad  j_{kl}/q_{kl}^*=\sum_{(i,j)\in{\cal G}_1}^{i,j\in{\cal G}_0}\frac{\partial h_i({\bm q}^*)}{\partial q_{kl}} (q_{ij}^*-1)\lambda_{ij}+
h_k({\bm q}^*)\lambda_{kl},
\label{SP}
\end{eqnarray}
where ${\cal S}({\bm j})$ is a convex function of its argument, also called the Cr\'{a}mer (or large-deviation) function. The dependence of the Cr\'{a}mer function on the current production ${\bm j}$ is defined implicitly via Eqs.~(\ref{SP}) and Eqs.~(\ref{local_i_q}).

Since $\Delta$ is fully defined by the solution of Eqs.~(\ref{local_i_q}), which is independent of ${\bm m}$, the resulting expression for the Cr\'{a}mer function is also ${\bm m}$-independent in the ``uncongested" regime.  This cancelation is quite remarkable.  We note that the degeneracy is especially interesting, since it does not seem to extend to the time-independent pre-factor in $P({\bm J}|t)$, $\Psi({\bm q})$ $\big($ defined in Eq.~(\ref{Psi_q}) $\big)$.  A Queueing Theory interpretation is that in this regime the large deviations are \textbf{not} caused by the interactions of different particles in the network, and thus \textbf{the same} kind of deviations would have occured even if all nodes in the network were of $M/M/\infty$ type (as opposed to $M/M/m$ with $m < \infty$), in which different particles cannot interact and delay one-another in the network.
  \footnote{Similar ``mysterious" cancelation of the vorticity dependence was reported in \cite{07TCCP} in the Cr\'{a}mer function of the entropy production for a polymer stretched by shear-vorticity flow.}.

As explained above, the qualitative assumption that allowed us to extend the product-form ansantz to statistics of currents in Eqs.~(\ref{global_q}-\ref{SP}) was that the number of particles in the system (size of the queue averaged over time) did not diverge with time.  For this assumption to hold, it must be the case that
\begin{eqnarray}
\forall i\in {\cal G}_0:\quad \langle \bar{n}_i\rangle_{\bm q}=\frac{
\langle {\bm 0} | \exp\left(\sum_{j\in{\cal G}_0}
\bar{h}_j({\bm q})\hat{a}_j\right) \hat{a}^+_i\hat{a}_i\prod_{k\in{\cal G}_0}
g_{m_k}(h_k({\bm q})\hat{a}^+)|{\bm 0}\rangle}{\langle {\bm 0} | \exp\left(\sum_{j\in{\cal G}_0}
\bar{h}_j({\bm q})\hat{a}_j\right) \prod_{k\in{\cal G}_0}
g_{m_k}(h_k({\bm q})\hat{a}^+)|{\bm 0}\rangle}<\infty,
\label{criterium}
\end{eqnarray}
where ${\bm h}({\bm q})$ and $\bar{\bm h}({\bm q})$ are solutions of Eqs.~(\ref{global_q}, \ref{local_i_q}) and Eqs.~(\ref{bar-eq1}, \ref{bar-eq2}) (which describe consistently the right/ket and left/bra ground state eigen-function of the evolution operator/Hamiltonian (\ref{Hq}) ).

We now comment on the phase transition which takes us from the uncongested to the congested regime.  Picking a direction in the vector space of currents (dimensionality of the space is equal to the number of internal directed edges of the graph) and increasing the length of the vector in this direction, gradually (starting from the domain of values belonging to the ``uncongested" regime) we will eventually reach the regime where congestion occurs, i.e. the condition (\ref{criterium}) breaks down and particles accumulate over time at some nodes (condensation).  To see this heuristically, we note that for large values of $q$, the main contribution to the generating function will be highly skewed towards very large values of current (due to the fact that $q$ is raised to a power equal to the current), and the most likely way to attain such large values of currents will be for the queueing network to remain totally occupied over the entire time horizon, causing the number of particles in the network to diverge over time. Put in the context of large deviations, the most likely way to attain very large values of currents is to have an accumulation over time of the number of particles in the network (condensation), leading to the system being overloaded over the entire time horizon, enabling all links in the network to generate current continuously over the entire time horizon.  We focus in particular on the setting where the ``breakdown" of the regime occurs initially at some particular unique node of the network, the expected number of particles in the system diverges at that node (over time), and a certain spectral condition holds at this breakdown point ( described in some detail below ). We will use these characteristics to formally define our ``congested" regime.

We now elaborate on the aforementioned spectral condition.  In particular, the transition point from the ``uncongested" to the ``congested" regime has an interpretation in terms of the closing of the gap between the ground state and the low boundary of the continuous spectrum of Eq.~(\ref{Hq}). This ``closing the gap" picture also assumes that other excited discrete states (which are present already in the case of a single station with feedback and $m>1$, see the discussion in Section \ref{sec:feedback}) do not cross the factorized ground state found above. This spectral interpretation also leads to an immediate conclusion/consequence in terms of the Cr\'{a}mer function shape at the values of the current that correspond to the considered uncongested-to-congested transition.  Indeed, this ``closing the gap" scenario translates into continuity of the Cr\'{a}mer function and its first derivatives at the transition. Stated differently (in the jargon of phase transition theory),
 the dynamical uncongested-to-congested transition with respect to the currents is second-order, and $1/\langle \bar{n}_i\rangle_{\bm q}$ plays the role of the order parameter at the congested node.

Our final remark is about comparison of the distribution of queue averaged over time with respective distribution measured at the final moment $t$. The asymptotic analysis, described above for uncongested regime and extended in the following Subsection to congested regime, suggests that $\langle n_i(t)\rangle_{\bm q}$ is finite at any value of $q$ and $t\to\infty$,  in particular for $q=q_c$ where $\langle \bar{n}_i\rangle_{\bm q}=\infty$. Moreover, one conjectures that the two distribution functions, $P_{\bm q}(\bar{n})$ and $P_{\bm q}(n(t))$, are different at any values of $q$ except of the special one correspondent to the minimum of the Cr\'{a}mer function achieved at ${\bm J}=\langle {\bar J}\rangle$. This statement may be interpreted as a breakdown of ergodicity for any current but the special one correspondent to the steady distribution.

\subsection{Congested Regime}
\label{subsec:congested}

\begin{figure}[b]
\includegraphics[width=3in,page=1]{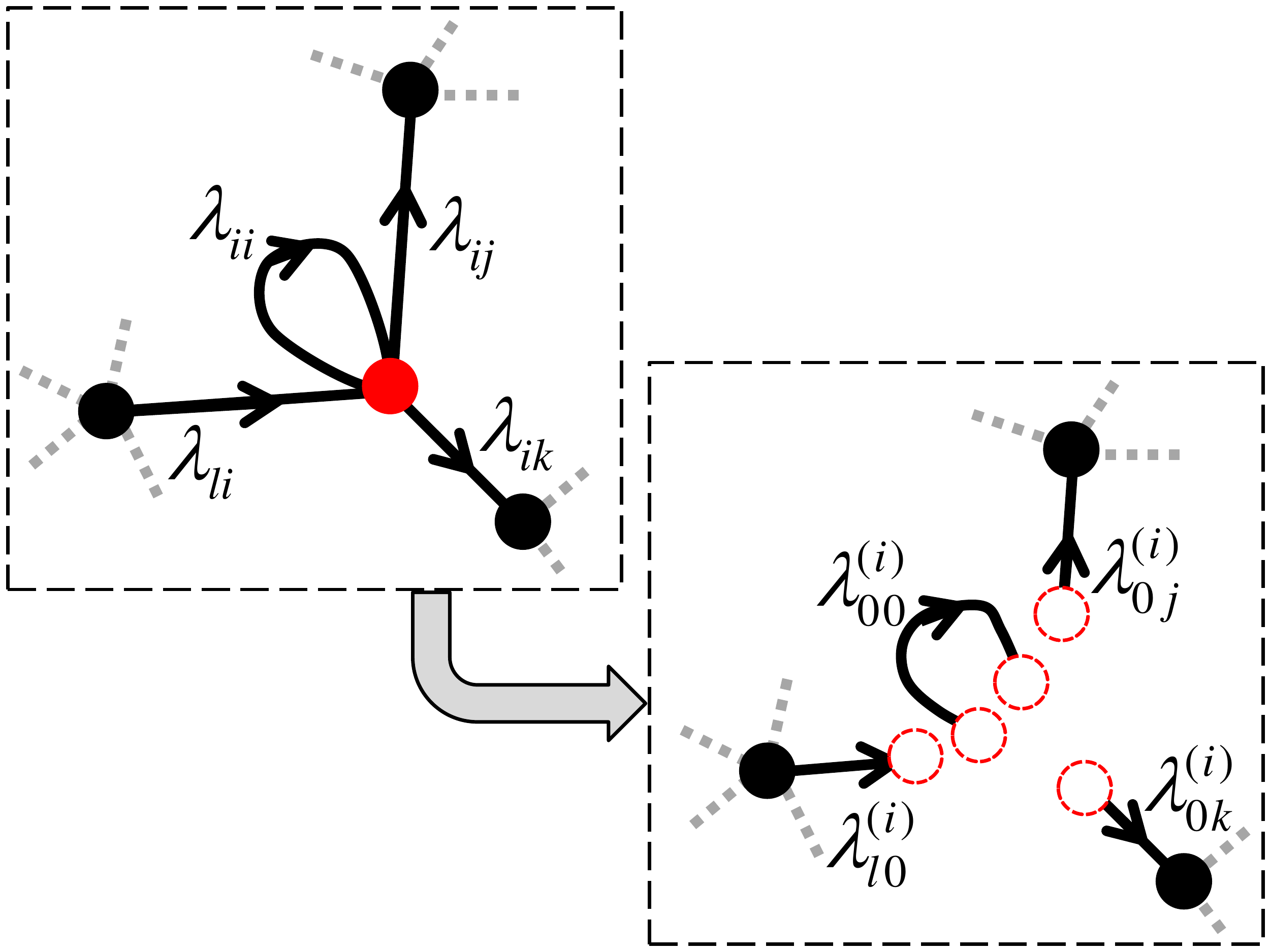}
 \caption{\label{fig:graph_modif}  Transformation of the ``overloaded" node,  discussed in the text. The component of the graph,  associated with the node, before and after the transformation are shown in the left-upper and low-right corners respectively.  Dashed node on the modified graph correspond to new injections and departures with the Poisson rates exactly equivalent to the respective transition rates on the original graph. }
 \end{figure}

We now consider a heuristic decomposition approach to extend the product-form description, utilized above in the uncongested regime, beyond the domain bounded by Eqs.~(\ref{criterium}).
We note that similar decomposition techniques have been considered throughout the  Queueing theory literature to understand how networks become congested \cite{M99,AFM09,DNR94,ES05,ES04,VCW94,D99,D00,BPT98}.

Consider,  for example, crossing the boundaries of Eq.~(\ref{criterium}) along a direction in the ${\bm q}$-space, and thus violating the condition at some ${\bm q}_*$ at a single node, say $i$. Then, exactly at the point of crossing we can simply assume that the node is always congested (has an infinite queue in the waiting room), and thus the tellers at the node stay busy over the entire time horizon. This translates into the following obvious modification of the network graph: remove the node, and associate all the in/out edges for the node of the old graph to the new open-system in/out nodes with exactly the same rates. The transformation is shown in Fig.~(\ref{fig:graph_modif}). In the new domain the Eqs.~(\ref{global_q}-\ref{local_i_q}) should be considered on the modified graph, and one needs to add the appropriate constant to $\Delta({\bm q})$ from Eq.~(\ref{Delta}) to guarantee its continuity at ${\bm q}_*$.  Equivalently, one must compensate all probabilities to reflect the rare event that the given node remained busy over the entire duration.

The general construction is illustrated in Fig.~(\ref{fig:feedback}),  and will also be illustrated  in the next Subsection on our enabling example of a single node system with feedback.
We note, of course, that in general the most likely way for rare events to occur in queueing networks may be quite complicated (see for example \cite{08MR}), and what was explained above should be considered as an approximation which is not proven rigorously.

We conclude this general construction by explaining the formal status of our derived results.  All the asymptotic derivations so far, discussed in both the congested and uncongested cases, have relied on several important assumptions.  The most important of these is the equivalence of the coherent state solution to the lowest eigen-value (ground state) solution.
We have assumed that this (physically very plausible) assumption holds, and focused primarily on constructing the coherent state solution explicitly.  We note, however, that our derivations have generally been non-rigorous in nature, and this assumption was not justified in any particular case.
A considerable difficulty associated with formalizing our results (and justifying this assumption) lie in the fact that the coherent state approach does not allow us to analyze the entire spectrum of the operator.  In view of the above, we find it important to validate this assumption, at least for a special case.  We now proceed along these lines by studying the aforementioned case of a single station with feedback.

\subsection{Single station with feedback}
\label{subsec:feedback}

\begin{figure}[t]
\includegraphics[width=3in,page=2]{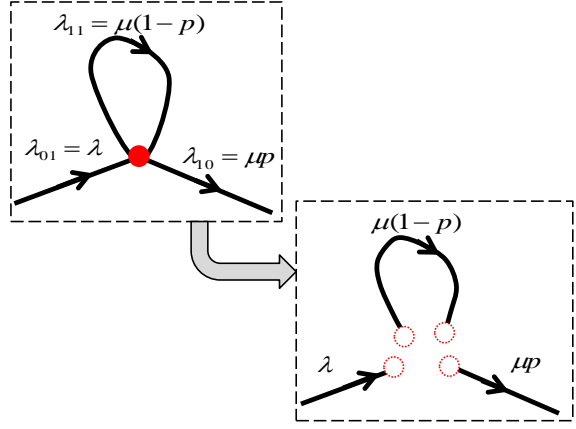}
 \caption{\label{fig:feedback}  Original graph and transformed graph correspondent to uncongested and congested regimes for example of the single station with the feedback.}
 \end{figure}

Perhaps the simplest example of a (non-trivial) queueing network (as discussed in Section \ref{feedbackground}) consists of a single station, one incoming channel/edge,  one outgoing channel/edge, and one self-loop, all characterized by Poisson processes with rates $\lambda$, $\mu p$ and $\mu (1-p)$ respectively. The network is shown in the upper left corner of Fig.~(\ref{fig:feedback}).

In this example, the only interesting current is associated with the self-loop (as the others are Poissonian in the steady-state, see Section\ \ref{feedbackground}).  We thus focus exclusively on the analysis of the current along the feedback arc, and the associated generating function (evaluated for the scalar $q$ and corresponding current $j$).

We start by considering the ``uncongested" regime. Then Eqs.~(\ref{global_q},\ref{Delta}) become
\begin{eqnarray}
h=\frac{\lambda}{\mu(1-(1-p)q)},\quad \Delta=-\frac{\lambda (q-1)(1-p)}{1-(1-p)q}.
\label{feedback1}
\end{eqnarray}
According to Eqs.~(\ref{P_J1}, \ref{SP}), this results in
\begin{eqnarray}
j<j_c:\quad{\cal S}(j)=\lambda(1-p/2)-\frac{\sqrt{p\lambda(4 j+p\lambda)}}{2}+
j\ln\left(\frac{2j+p\lambda-\sqrt{p\lambda(4 j+p\lambda)}}{2j(1-p)}\right).
\label{feedbackCramer1}
\end{eqnarray}

\begin{figure}[b]
\includegraphics[width=3in]{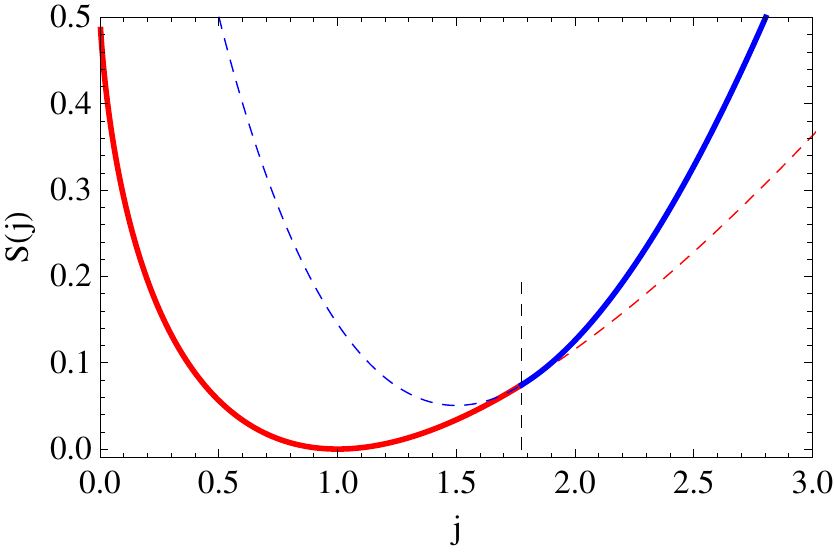}
 \caption{\label{fig:Cramer}
 Cr\'{a}mer function, ${\cal S}(j)$, of the feedback current shown for
 $\lambda=m=1$, $p=1/2$ and $\mu=3$. Thick red and thick blue curves describe the uncongested and congested domains respectively. Dashed black line marks the value of $j_c$.}
 \end{figure}

To identify the ``uncongested" regime breakdown we calculate $\bar{h}$,  according to Eq.~(\ref{bar-eq2}):
\begin{eqnarray}
\bar{h}=\frac{p}{1-(1-p)q}.
\label{feedback1-barh}
\end{eqnarray}
Substituting into Eqs.~(\ref{Psi_q}) results in
\begin{eqnarray}
&& \Psi(q)=g_m(\bar{h}(q)h(q)),\quad \langle \bar{n}\rangle_q=\frac{\partial}{\partial z}\left.\frac{g_m(\bar{h}(q)h(q)z)}{g_m(\bar{h}(q)h(q))}\right|_{z=1}.
\label{Psi-feedback}
\end{eqnarray}
From the above and Eq.~(\ref{cohm2}), we find that the expected queue length remains bounded over time iff $h\bar{h}<m$, and thus the critical value of congestion at which point one shifts regimes is
\begin{eqnarray}
q_c=\frac{1-\sqrt{\frac{\lambda p}{m\mu}}}{1-p}.
\label{qc-feedback}
\end{eqnarray}
The congested regime that occurs at $q>q_c$ and $j>j_c$ corresponds to the situation when the single server is always occupied, and thus (conditioning on this event) jobs continually feedback over time as a $m\mu(1-p)$ standard Poisson process. Therefore, in the congested regime
\begin{eqnarray}
q>q_c:\quad\Delta=-\mu(1-p)m(q-1)+\left(\sqrt{\lambda}-\sqrt{p\mu m}\right)^2.
\label{Delta_satur}
\end{eqnarray}
Here the $\left(\sqrt{\lambda}-\sqrt{p\mu m}\right)^2$ term (a constant with respect to $q$) is found in accordance with Eq.~(\ref{feedback1}) and the condition of $\Delta$-continuity at $q=q_c$. Alternatively, in the congested regime $\Delta$ can be identified with the lower edge $s_{-}$ of the continuous spectrum of $\hat{H}_{q}$, given by Eq.~(\ref{spectr-edges}). The spectrum of $\hat{H}_{q}$ for a simple model under consideration is analyzed in some detail in Section~\ref{sec:feedback}. Performing the Legendre transform on Eq.~(\ref{Delta_satur}), we arrive at the following expression for the Cr\'{a}mer function of the feedback current in the congested regime
\begin{eqnarray}
j>j_c:\quad {\cal S}(j)=\left(\sqrt{\lambda}-\sqrt{p\mu m}\right)^{2}+m\mu(1-p)
+j\ln(\frac{j}{em\mu(1-p)}).
\label{Cramer_satur}
\end{eqnarray}
Comparing Eq.~(\ref{feedbackCramer1}) with Eq.~(\ref{Cramer_satur}) we also observe that the Cr\'{a}mer function is smooth (first derivative is continuous) at $j=j_c$, thus confirming that the dynamical phase transition described here is of the second order (continuous). The change of the Cr\'{a}mer function shape across the transition is shown in Fig.~\ref{fig:Cramer}.

The obtained expression for the Cr\'{a}mer function in the congested regime Eq.~(\ref{Cramer_satur}) has a very simple and transparent interpretation. Since the server is always occupied, the feedback current generation is a standard Poisson process, so that ${\cal S}(j)=\mu m(1-p)+j\ln(j/(em\mu(1-p)))$, where $e^{-t{\cal S}_{0}}$ is the probability to keep the server busy, presented with exponential accuracy. The latter is given by the probability $e^{-t{\cal S}_{0}(\omega)}$ of creating the incoming and outgoing currents of the same value $\omega$ to provide the marginal stability of the server, maximized with respect to $\omega$. Since both currents are generated by independent Poisson processes with the rates $\lambda$ and $p\mu m$, respectively, we have ${\cal S}_{0}(\omega)=\lambda+pm\mu+\omega\ln(\omega^{2}/(\lambda p\mu m)e^{2})$, and minimization with respect to $\omega$ results in ${\cal S}_{0}=(\sqrt{\lambda}-\sqrt{p\mu m})^{2}$, which reproduces Eq.~(\ref{Cramer_satur}).

\section{Direct Analysis of the singe-station feedback system}
\label{sec:feedback}

As shown in the previous Sections, an understanding of the evolution operator ground state structure is essential for predicting long-time statistics of currents.  We have also argued that the ground state can be described analytically for networks with general graphical structure.  However, these arguments were indirectly dependent on certain information about the rest of the spectrum - specifically on the fact that the ground state is separated from the continuous spectrum by a gap which collapses at the dynamical phase transition.  In general, information on the entire spectrum is difficult to obtain.  The main point of this Section is to gain a broader understanding of the simple single-node network with feedback along these lines.  Therefore in this Subsection we analyze the full spectrum of the problem with the single feedback and we confirm the general picture of the `gap emergence', and collapse at the phase transition point suggested above on the basis of only partial (ground state) analysis.

Our starting point is the dynamical equation $P_q(n;t)=\sum_{j=0}^\infty q^j P(n,j;t)$ for the generating function of current $j$ via the feedback arc. Here $P(n,j;t)$ is the joint probability distribution function of the number of particles, $n$ standing at the queue at the time $t$, and the number of particles, $j$, that have passed through the feedback loop by time $t$.  Following directly the proper generalization, according to Eqs.~(\ref{current}-\ref{Hq}) of the ME~(\ref{MEm}) we derive:
\begin{eqnarray}
&& \frac{\partial}{\partial t} P_q(n;t)=\lambda\left(P_q(n-1;t)-P_q(n;t)\right)+
\mu p \left(\theta_m(n+1)P_q(n+1;t)-\theta_m(n)P_q(n;t)\right)\nonumber\\
&& +\mu (1-p)(q-1)\theta_m(n)P_q(n;t),
\label{feedbackS1}
\end{eqnarray}
where the last term on the r.h.s. accounts for the current.
Performing the Laplace transform of $P_q(n;t)$ over time,
$P_{s;q}(n)\equiv\int_0^\infty \exp(s t) P_q(n;t) dt$ (with $s$ considered as a spectral parameter), we arrive at the following spectral equation:
\begin{eqnarray} \label{Pseq}
 (s-\lambda-\mu p\theta_m(n)+\mu(1-p)(q-1)\theta_m(n)) P_{s;q}(n) = - \lambda P_{s;q}(n-1) - \mu p \theta_m(n+1) P_{s;q}(n+1).
\end{eqnarray}
Here the inhomogeneous part, dependent on the initial condition at $t=0$, is ignored.

The ``boundary" conditions over $n$ for Eq.~(\ref{Pseq}) read  $P_{s;q}(-1) = 0$ and $P_{s;q}(n\to\infty) =0$.  We will relax the last condition, assuming the following natural finite waiting room regularization: $P_{s;q}(N+m) = 0$ for some sufficiently large value of $N$.  We will study the spectrum at finite $N$, and show towards the end of the calculations that the $N\to\infty$ limit is well defined.  We note that the particular choice of regularization turns out to be unessential for the limits we consider.  In order to find the eigenvalues (spectrum) we solve the recurrence relation (\ref{Pseq}) for general values of $s$, and use the aforementioned boundary condition, and the normalization condition $P_{s;q}(0)=1$.  To summarize, the set of conditions that complement Eq.~(\ref{Pseq}) to provide the full description of the spectrum are
\begin{equation}
P_{s;q}(-1) =0, \quad P_{s;q}(0)=1, \quad P_{s;q}(N+m) = 0,
\label{conds}
\end{equation}
and we are mainly interested in studying the $N\to\infty$ limit.

\begin{figure}[t]
\includegraphics[width=5in]{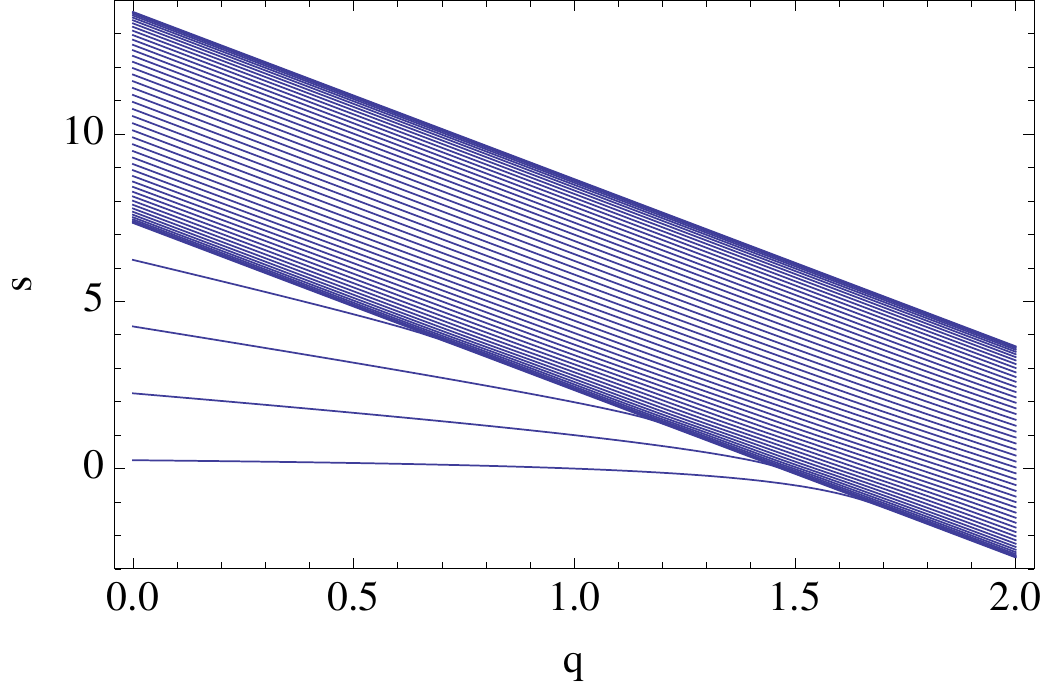}
 \caption{\label{fig:Spectrum}
Spectrum of the evolution operator as a function of $q$ for $m=5,\lambda = 0.5,\mu = 2.0, p=0.5$ and $N=50$. The  lower bottom line corresponds to the ground state solution $\Delta(q)$.}
\end{figure}

A numerical solution of the eigenvalue problem is illustrated in Fig.~(\ref{fig:Spectrum}) for
$m=5,\lambda = 0.5,\mu = 2.0, p=0.5$ and $N=50$. The simulations show emergence at $q=1$ of three discrete eigenstates well separated from the (still discrete) band of states. We tested numerically as one increases $N$, the $q=0$ value of the ground state $s$ (top line in Fig.~\ref{fig:Spectrum}) approaches $0$.  We observed that the asymptotic dependence of the ground state eigen-value on $s$ is fully consistent with the prediction of the ground-state coherent-state theory explained above in Section \ref{subsec:feedback}, specifically with
Eq.~(\ref{feedback1}).  Moreover, our prediction of $q_c$ (where the gap between the ground state and the continuous band collapses), as given by Eq.~(\ref{qc-feedback}), is fully consistent with the respective dependencies of the gap collapse in our computations.  Let us also mention that the two non-ground state eigen-values observed at $q=1$ never cross with each other or with the ground state at $q>1$, and merge into the continuous band (consequently one after another) at values of $q$ smaller than $q_c$. Experimenting with larger $m$, we observe that the discrete spectrum becomes equidistant in the $m\to\infty$ limit, which is fully consistent with the interaction-free nature of the limit.

We now analyze the spectrum analytically.  Let us begin by making some preliminary observations.
First, the linear dependence on $s$ implies that the function $P_{s;q}(n)$ is an $n$-th order polynomial in $s$, suggesting that there are always exactly $N+m$ solutions of the last condition in Eq.~(\ref{conds}), $P_{s;q}(N+m) = 0$. Second, at $n>m$ the factor $\theta_m(n)$ becomes constant and the recursive relations (\ref{Pseq}) can be solved analytically. Thus, looking for solution of Eqs.~(\ref{Pseq}) at $n>m$ in the $P_{s;q}(n) = c_+ \rho_+^n + c_- \rho_-^n$ form, one obtains the following expression for $\rho_\pm$ (which depends on $q,s$ and the rates):
\begin{equation}
 \rho_\pm = \frac{1}{2\mu p m}\left(\lambda + \mu p m - \mu (1-p) (q-1)m-s \pm \sqrt{(\lambda + \mu p m - \mu (1-p) (q-1)m -s)^2 - 4 \lambda \mu p m}\right).
 \label{rhopm}
\end{equation}
The boundary condition $P_{s;q}(N+m) = 0$ translates into the following condition $\big($ dependent on $P_{s;q}(m)$ and $P_{s;q}(m-1) \big)$:
\begin{equation}\label{BCeq}
P_{s;q}(N+m) = \frac{(P_{s;q}(m)-\rho_- P_{s;q}(m-1))\rho_+^N + (P_{s;q}(m)-\rho_+ P_{s;q}(m-1))\rho_-^N}{\rho_+-\rho_-} = 0.
\end{equation}
The above has different types of solutions dependent on the values of $s$ with respect to the following two threshold values:
\begin{eqnarray}
\label{spectr-edges} s_\pm = \lambda + \mu p m + \mu (1-p)(q-1)m \pm\sqrt{4 \lambda \mu p m}= \left(\sqrt{\lambda} \pm \sqrt{\mu p m}\right)^{2} - \mu (1-p)(q-1)m.
\end{eqnarray}
At $s< s_-$, both eigenvalues $\rho_\pm$ are real and $\rho_+>\rho_-$. In this case, as we are interested in the $N\to\infty$ limit, one can simply ignore the $\rho_-^N$ contributions in (\ref{BCeq}), thus leading to the following relation:
$P_{s;q}(m) - \rho_-P_{s;q}(m-1) = 0$.  This then replaces
the last condition in Eq.~(\ref{conds}), w.r.t. describing $P_{s;q}(n)$ at $0\leq n\leq m$.
We were not able to solve this reduced system of equations analytically at any values of $m$, and thus to find the spectrum in its full glory.
However, the information just provided is already sufficient for a heuristic derivation of the lowest eigenvalue, and the value of the gap between it and the continuous band.

We focus on a particular single form solution of the reduced system of equations for $P_{s;q}(n)$ with $0\leq n\leq m$, corresponding to $s = \Delta $ and $P_{s;q}(n) = h^n/n!$ for $n\leq m$, with $h$ and $\Delta$ from Eq.~(\ref{feedback1}).  In this case, one has $\rho_+ = \lambda/(\mu p h)$ and $\rho_-= h/m$, so this is indeed the ground state solution.
However, let us note for the sake of accurateness that we have not formally proven that this special solution always corresponds to the lowest eigenvalue (the ground state).  However, this is exactly what we observed in our numerical experiments with different values of $m$.

Eq. (\ref{BCeq}) also allows the continuous spectrum of the operator to be identified. At $s_- < s < s_+$ we have $|\rho_+| = |\rho_-|$, and one has to keep both terms in Eq.~(\ref{BCeq}). As long as $|P_{s;q}(m+1)-\rho_- P_{s;q}(m)| = |P_{s;q}(m)-\rho_+ P_{s;q}(m-1)|$, Eq.~(\ref{BCeq}) has at least $N$ solutions. At $N\gg 1$ this region turns into the continuous band of the spectrum. We conjecture that the system of equations does not have any solutions for $s>s_+$. (Once again, this is confirmed in simulations but we do not have an explicit way of proving it.)

Finally, we conclude that the transition between the uncongested regime and congested regime takes place at $\Delta = s_-$ which corresponds exactly to $q=q_c$ $\big($ from Eq.~(\ref{qc-feedback}) $\big)$. As one can easily see, the value of $h$ at this point is equal to $h = \sqrt{\lambda/\mu p} < 1$, so the queue length in the final moment does not diverge.

\section{Conclusions and Path Forward}
\label{sec:path}

Let us briefly recall the highlights of this manuscript. Our analysis was focused on the the generating function of currents over a Jackson (queueing) network, with an eye towards analyzing how large currents accumulate over time.  We began by giving some relevant background in Queueing Theory, and discussing some tie-ins with recent work in statistical physics.  We then adopted the Doi-Peliti technique for describing the dynamics of the Jackson network.  We used this formalism to show that the ground state of the respective evolution operator has a well-defined and analytically tractable product/coherent state form in a particular regime.
These results were translated into an implicit analytical expression for the Cr\'{a}mer function of currents in this ``uncongested" regime,
where the ground state was well separated by a gap from the continuous spectrum.  We also observed that crossing the surface in the phase space of observed currents, where the spectral gap collapses, corresponds to the congestion of a node in the network.  We suggested a heuristic graph-reduction scheme which allows the statistics of currents to be described in this partially congested setting. Finally, we validated the general results using the example of a single node feedback system, where many of our assumptions could be verified directly by performing an explicit analysis of the evolution operator spectrum.

We consider this study more like an opening for further exciting research along the following lines:
\begin{itemize}
\item Following the discussion in Section \ref{subsec:congested}, we naturally conjecture that in a large network, gradually increasing the observed current(s) will lead to a transition from the uncongested to the fully congested regime via a number of steps, each characterized by an increase in the number of congested nodes. It would be interesting and important to explore the specific sequence of phase transitions separating the space of observed currents into cells.  A particularly interesting question concerns the algorithmic complexity of identifying these cells and exploring their geometric structure (e.g., possible convexity).
\item We have considered a queuing model with the transition rates independent of the node occupation numbers. This condition can be relaxed and such an extension of our theory should allow for non-uniform dependence of the rates on the occupation numbers.
\item Extensions of the current-statistics theory to the so-called multi-class networks, where different classes of particles are treated differently at the stations (for example having different priorities) are less trivial, yet we anticipate such a generalization is still possible.
\item In this manuscript we have considered solely open queuing networks.  It would be interesting and instructive to generalize the theory of current statistics to the case of closed (particles are not leaving or entering the network) and semi-open (particles are injected into the system and leaving it, but in such a way that the total number of particles is conserved) networks.
\item We may also consider networks of fixed structure, yet with rates changing in time, for example in a periodic fashion.  To describe statistics of currents in this case,  and especially in the regime where the typical correlation time of the rate changes are comparable to the inverse rates, constitutes another interesting future challenge. (Note that an approach blending the techniques described in this manuscript with the ones discussed recently in the context of the so-called Jarzynski equality \cite{97Jar} and work relations \cite{99Cro,05CCJ}, both closely related to the subject of fluctuation theorems \cite{95GC,98Kur,99LS}, may prove fruitful for this task.)
\item The assumption of infinite waiting room was crucially important for advancing the product/coherent state decomposition to the statistics of currents. Analyzing the current statistics in the regimes when all or some waiting rooms are of a finite capacity is yet another challenging task, as this finiteness brings in a new type of inter-particle interaction.
\item It would be an interesting challenge to put the ideas presented in this paper on a more rigorous mathematical foundation. This would aid greatly in understanding the formal relationship between the regimes identified in this paper and the sample path large deviations properties of queueing networks.  We note that the authors are currently undertaking preliminary work along these lines, with the work directed more towards the Queueing Theory community.
\item Our analysis suggests that in large queuing networks one may expect emergence of multiple transitions as one increases the current. More generally, and in the spirit of \cite{89KKR,90DKRS}, it would be interesting to explore the field of the so-called qualitative queuing network theory via the methods/techniques discussed in this manuscript.
\item Finally, all of the above should be used not only to study existing (man- or nature- made) networks, but also to
    guide construction of future technological networks with desired properties.  In other words, we suggest to use this analysis for control and optimization of networks in new areas such as power, and even more generally energy, distribution.
\end{itemize}

\section{Acknowledgments}

We are thankful to David Gamarnik for consulting us on many issues related to Queuing Theory, and Sergey Foss, Bill Massey and Alexander Rybko for enlightening conversations. This material is based upon work supported by the National Science Foundation under CHE-0808910 (VC) and CCF-0829945 (MC via NMC). The work at LANL was carried out under the auspices of the National Nuclear Security Administration of the U.S. DoE at LANL under Contract No. DE-AC52-06NA25396. KT acknowledges support of an Oppenheimer Fellowship at LANL, and DAG work on the project was a part of his summer internship (GRA program) at LANL.

\appendix

\section{Left Ground State of the Hamiltonian}
\label{app:Left}

In this Appendix we construct the left ground state of the Hamiltonian (\ref{Hq}). We start by recalling that according to Eq.~(\ref{constr}), the bra-vector $\langle {\bm 0}|\exp(\sum_i \hat{a}_i)$ is the left zero eigen-function of the Hamiltonian (\ref{Hq}) at ${\bm q}={\bm 1}$.  However, it ceases to be an eigen-vector at ${\bm q}\neq {\bm 1}$.  On the other hand, it is easy to check that the ``exponential" bra-vector $\langle 0|\exp(\bar{h}\hat{a})$ is in fact a left eigen-vector of the creation operator $\hat{a}^+$, with the eigen-value $\overline{h}$, $\langle 0|\exp(\overline{h}\hat{a})\hat{a}^+=\overline{h}\langle 0|\exp(\bar{h}\hat{a})$. This suggests searching for a left eigen-vector of the Hamiltonian (\ref{Hq}),
in the exponential form:
\begin{eqnarray}
\langle {\bm s}_{\bm q}|=\langle {\bm 0}|\exp\left(\sum_i \bar{h}_i\hat{a}_i\right).
\label{left}
\end{eqnarray}
Then, from $\langle {\bm s}_{\bm q}|\hat{H}_{\bm q}=-\bar{\Delta}({\bm q})\langle {\bm s}_{\bm q}|$ and Eq.~(\ref{Hq}) (combined with utilizing the aforementioned feature of the creation operators) one derives the following set of conditions on the $\bar{\bm h}$ vector (of $c$-numbers):
\begin{eqnarray}
&& \sum_i (q_{0i}-\bar{h}_i({\bm q}))\lambda_{0i}=\bar{\Delta}({\bm q})
\label{bar-eq1}\\
&&
\forall i:\quad \lambda_{i0}\left(q_{i0}-\bar{h}_i({\bm q})\right)+
\sum_{j}^{(i,j)\in {\cal G}_1}\lambda_{ij}(q_{ij}\bar{h}_j({\bm q})-\bar{h}_i({\bm q}))=0.
\label{bar-eq2}
\end{eqnarray}
It is straightforward to verify that the resulting $\bar{\Delta}({\bm q})$ and $\Delta({\bm q})$ from
Eq.~(\ref{global_q}) are identical, $\bar{\Delta}({\bm q})=\Delta({\bm q})$.

\bibliographystyle{apsrev}
\bibliography{queuing}

\end{document}